\documentclass[12pt,preprint]{aastex}
\usepackage{amsfonts,amsmath,graphicx,natbib}

\shortauthors{Bower et al.}
\shorttitle{Late Time Radio Emission from TDEs}
\begin{document}

\newcommand{\ubslong}{Swift J164449.3+573451}
\newcommand{\ubs}{Sw 1644+57}

\newcommand\degd{\ifmmode^{\circ}\!\!\!.\,\else$^{\circ}\!\!\!.\,$\fi}
\newcommand{\etal}{{\it et al.\ }}
\newcommand{\uv}{(u,v)}
\newcommand{\rdm}{{\rm\ rad\ m^{-2}}}
\newcommand{\msuny}{{\rm\ M_{\sun}\ y^{-1}}}
\newcommand{\mylesssim}{\stackrel{\scriptstyle <}{\scriptstyle \sim}}
\newcommand{\lsim}{\stackrel{\scriptstyle <}{\scriptstyle \sim}}
\newcommand{\gsim}{\stackrel{\scriptstyle >}{\scriptstyle \sim}}
\newcommand{\sci}{Science}
\newcommand{\boo}{Bo\"{o}tes\ }

\def\kbar{{\mathchar'26\mkern-9mu k}}
\def\totd{{\mathrm{d}}}
\newcommand{\be}{\begin{equation}}
\newcommand{\ee}{\end{equation}}


\title{Late Time Radio Emission from X-ray Selected Tidal Disruption Events}

\author{
 Geoffrey C.\ Bower\altaffilmark{1},
Brian D.~Metzger\altaffilmark{2}, 
S. Bradley Cenko\altaffilmark{1}, 
Jeffrey M. Silverman\altaffilmark{1},
Joshua S. Bloom\altaffilmark{1}
}
\altaffiltext{1}{Astronomy Department and Radio Astronomy Laboratory, University of California, Berkeley, 601 Campbell Hall \#3411, Berkeley, CA 94720, USA; gbower@astro.berkeley.edu }
\altaffiltext{2}{Department of Astrophysical Sciences, Peyton Hall, Princeton University, Princeton, NJ 08542, USA}

\begin{abstract}
We present new observations with the Karl G. Jansky Very Large Array of seven X-ray-selected tidal disruption events (TDEs).  The radio observations were carried out between 9 and 22 years after the initial X-ray discovery, and, thus, probe the late-time formation of relativistic jets and jet interactions with the interstellar medium in these systems.  We detect a compact radio source in the nucleus of the galaxy IC 3599 and a compact radio source that is a possible counterpart to RX J1420.4+5334.  We find no radio counterparts for five other sources with flux density upper limits between 51 and 200 $\mu$Jy ($3\sigma$).  If the detections truly represent late radio emission associated with a TDE, then our results suggest that a fraction $\gsim 10$\% of 
X-ray-detected TDEs are accompanied by relativistic jets.  We explore several models for producing late radio emission, including interaction of the jet with gas in the circumnuclear environment (blast wave model), and emission from the core of the jet itself.  Upper limits on the radio flux density from archival observations suggest that the jet formation may have been delayed for years after the TDE, possibly triggered by the accretion rate dropping below a critical threshold of $\sim 10^{-2}$ -- $10^{-3} \dot{M}_{Edd}$.  The non-detections are also consistent with this scenario; deeper radio observations can determine whether relativistic jets are present in these systems.  The emission from RX J1420.4+5334 is also consistent with the predictions of the blast wave model, however the radio emission from IC 3599 is substantially underluminous, and its spectral slope is too flat, relative to the blast wave model expectations.  Future radio monitoring of IC 3599 and RX J1420.4+5334 will help to better constrain the nature of the jets in these systems.    

\end{abstract}


\section{Introduction}
\label{sec:intro}

A star that passes too close to a supermassive black hole (SMBH) may be disrupted by tidal forces, leading to a transient accretion event.  Such a ``tidal disruption event (TDE)'' is predicted to be visible at optical, ultraviolet, and X-ray wavelengths \citep[e.g.,][]{1988Natur.333..523R,1999ApJ...514..180U,2009MNRAS.400.2070S}.  In the simplest scenario, the process of stellar disruption leaves $\sim 1/2$ of the star bound to the SMBH, resulting in a mass accretion rate $\dot{M}$ that can initially exceed the Eddington limit, before declining as a power-law $\dot{M} \propto t^{-5/3}$ with a characteristic time scale of weeks to months.  Since the inner accretion disk reaches temperatures $\gsim 10^5$ K, this leads to a luminous thermal flare peaking at UV/X-ray wavelengths.

Over the past two decades, there have been numerous claimed observational detections of the TDE phenomenon.  The earliest TDE candidates were detected by the ROSAT X-ray telescope \citep[][and references therein]{1999A&A...343..775K,2002AJ....124.1308D}.  More recently, a number of events have been detected at ultraviolet \citep{1995Natur.378...39R,2006ApJ...653L..25G,2012Natur.485..217G} and optical 
\citep{2011ApJ...741...73V,2012MNRAS.420.2684C}
wavelengths, in addition to recent X-ray flare candidates (e.g., \citealt{2012A&A...541A.106S}).  Several well-studied events have shown the characteristic impulsive flare, followed by a 
light curve decline, consistent with the power-law that is expected following a TDE.

Accretion and outflows are coupled in a wide range of astrophysical systems, from proto-stars to stellar mass black holes to the billion solar mass black holes in galactic nuclei.   However, the mechanism by which relativistic jets are launched remains poorly understood.  TDEs represent a dynamic accretion system which can in principle provide a unique testbed to explore what conditions are required for jet creation and how the jet properties change as $\dot{M}$ evolves.  The characteristic duration of jet activity $t_{\rm j}$ in a TDE, for instance, can vary substantially depending on the theoretical model.  If the jet luminosity faithfully tracks the accretion rate at all times, then the jet duration is similar to the characteristic fall-back time of the most bound stellar debris, which can vary from days to months depending on the mass of the SMBH and the pericenter distance of the stellar orbit (e.g.~\citealt{2009MNRAS.400.2070S}, \citealt{2012arXiv1206.2350G}).  The jets in galactic microquasars, on the other hand, are suppressed when $\dot{M}$ is high, but then appear when $\dot{M}$ drops below a critical threshold $\sim 10^{-3}-10^{-2}\dot{M}_{\rm edd}$  \citep{2004ARA&A..42..317F}.  If a similar ``state transition'' applies in the case of TDEs, then jet creation could be delayed for years or longer. 

Two models have been proposed for radio emission from relativistic jets in TDEs.  \citet[][hereafter GM11]{2011MNRAS.416.2102G} show that radio synchrotron emission can originate from the shock interaction between the transient relativistic jet and the dense gas surrounding the SMBH (the `circumnuclear medium', or CNM).  Emission during the earliest phases of the jet-CNM interaction occurs from both the forward shock plowing into the CNM and the reverse shock propagating back through the relativistic ejecta, with their relative contirbution to the emission output depending on the jet duration and CNM density.  Due to relativistic beaming, the radio emission, as observed from a typical location off the axis of the jet, peaks only once the blastwave decelerates to mildly relativistic speeds on a timescale $t = t_{\rm dec} \sim$ yr.  If the jet contains $\sim 1\%$ of the rest mass energy of the accreted star, then the peak radio luminosity on this timescale is approximately $L_{R} \sim 10^{30}$ erg s$^{-1}$ Hz$^{-1}$ and hence can be detected to large distances.  After the peak ($t \gtrsim t_{\rm dec}$), the blast wave soon enters a Sedov-Taylor expansion and the radio flux decreases as a power law $F_{\nu} \propto t^{-\alpha}$ with $\alpha \sim $ 1 -- 2.  A second model for radio emission from TDEs was proposed by \citet{2011MNRAS.417L..51V} based on emission internal to the jet, which they calibrate using observed scaling laws between accretion power and jet luminosity for radio loud sources.  The model predicts peak radio luminosities ranging from $10^{28}$ to $10^{31}$ erg s$^{-1}$ Hz$^{-1}$, depending on when the source becomes radio loud as a function of $\dot{M}$.  The timescales for peak radio luminosity range from $\sim 1$ to $\gtrsim 10$ years.  

Until recently, radio emission had not been discovered from a TDE.  \citet{2011ApJ...732L..12B} summarized the limited constraints from radio observations.  However, recently two TDEs first detected in the $\gamma$- and X-rays by the Swift satellite demonstrated the existence of bright and highly variable  radio emission on timescales of days to months following the TDE \citep{Bloom08072011,Levan08072011,2011Natur.476..425Z,2012ApJ...753...77C}.  For both objects, these studies found evidence for a mildly relativistic outflow beamed towards the Earth, jet collimation, and a spectrum characterized by synchrotron and inverse Compton processes, leading to a natural analogy of these objects  with a blazar.  
\citet[][hereafter MGM12]{2012MNRAS.420.3528M}
showed that the early radio emission from Swift J1644+57 could be well-explained by the CNM interaction model.  
\citet{2012ApJ...748...36B} model the jet/CNM interaction based on emission $\sim 1$ year after the TDE and find that 
the radio emission associated with the jet is likely to persist for decades, or longer.

One question raised by the discovery of the Swift TDEs is whether such energetic relativistic jets are a universal component of all TDEs, or whether their production requires special conditions, such as a rapidly spinning SMBH or highly super-Eddington accretion \citep[e.g.,][]{2011ApJ...738L..13M,2012ApJ...749...92K,2012ApJ...760..103D}.  In this paper, we begin to address this question by pursuing radio observations with the Very Large Array of a sample of X-ray detected TDEs.  We describe observations, analysis, and results in \S~\ref{sec:obs}, discuss our conclusions in \S~\ref{sec:discussion}, and summarize in \S~\ref{sec:conclusions}.

\section{Observations, Analysis, and Results \label{sec:obs}}

Our sample of objects include the five sources visible to the VLA
from systematic surveys of the ROSAT All Sky Survey
\citep[][, and references in the table]{2002AJ....124.1308D},
and two events discovered through XMM and Chandra observations.
In Table~\ref{tab:sample},
we list target X-ray data.
Columns are (1) source name, (2) discovery epoch of the X-ray transient, (3) distance to nominal host galaxy,
(4), peak X-ray luminosity, $L_X^{peak}$, (5) X-ray observatory used for discovery, and 
(6) references for discovery and the distance to the object.  Note that for sources
discovered with XMM or Chandra, the subarcsecond positions lead to a clear
 identification with a host galaxy and, therefore, the
distance estimate is relatively unambiguous.  ROSAT identifications with host galaxies
are less certain due to the larger error circles.  
We list here the likely association but note the discussion for
RX J1420.4+5334 that we introduce in later sections.  That likely association is often
determined by the most luminous optical galaxy in the field.
The ROSAT sources have characteristic ages at the present epoch of 22 years.  The XMM and Chandra
events have a characteristic age of $\lsim 10$ years.  Note that the discovery epoch for many of
these events is not well constrained due to the sparse coverage of X-ray observations.

The Very Large Array (VLA) observed seven TDE candidates on 04 and 06 June 2012.  The VLA was
in B configuration.  Observations were made in S band with a total of 2 GHz of bandwidth
per polarization over the frequency range 2 to 4 GHz.  The data were divided into 16 
subbands per polarization with 64 channels in each subband.  Each TDE source was paired with
a nearby phase calibrator and observed for approximately 15 minutes.  The bright calibrator
3C 286 was observed on both dates to set the absolute flux density scale.  

Data reduction was carried out with the CASA package, following standard interferometric
techniques.  Manual flagging of broad- and narrow-band radio frequency interference (RFI) was 
carried out.  The ability to fully flag RFI ultimately limited sensitivity and dynamic
range in the target images.  Images were made using multi-frequency synthesis techniques,
creating a mean flux density image and a first-order derivative image.  Images typically
had a resolution of 3\arcsec\ and an rms noise of $<20\, \mu$Jy.  In Table~\ref{tab:results} we summarize image statistics, and TDE counterpart detection data.  
Columns are (1) source name, 
(2) VLA beam size, (3) VLA image rms, $\sigma$,
(4) VLA target flux density, $S_{rad}$, (5) VLA target right ascension, and (6) VLA target declination.
For non-detections, we give a $3\sigma$ upper limit to the flux density.
Errors in position are $\lsim 1$\arcsec.  The flux density that we report is at the mean frequency
of the observations, 3.0 GHz.

\begin{deluxetable}{lrrrrl}
\scriptsize
\tablecaption{TDE Candidate X-ray Data \label{tab:sample}}
\tablehead{
\colhead{Source}&  
\colhead{Epoch}  &  
\colhead{Dist}  &  
\colhead{$L_X^{peak}$}  &  
\colhead{Obs.} &
\colhead{Ref.} \\ 
                           &  (year)  &  (Mpc)  &  ($10^{43} $ erg s$^{-1} $)   &   
}
\startdata
IC 3599                  & 1990.94  &   88 & 2.8   & ROSAT & 1,2 \\
RX J1420+5334            & 1990.94  &  2970$^{1}$ & 53.6    & ROSAT & 3\\
NGC 5905                 & 1990.52  &   52 & 0.05    & ROSAT & 4 \\
RX J1624+7554            & 1990.76  &  265 & 6.2    & ROSAT & 5\\
RX J1242-1119            & 1992.54  &  208 & 36   & ROSAT & 6\\
SDSS J132341.97+482701.3 & 2003.92  &  365 & 4.4    & XMM & 7,8 \\
SDSS J131122.15-012345.6 & 2004.16  &  750 & 0.5    & XMM/CHANDRA & 9 \\
\enddata

$^{1}$Assuming the true host galaxy is SDSS J142025.18+533354.9 (galaxy B) rather than the
galaxy identified in the discovery paper.
References:  
(1) \citet{1995MNRAS.273L..47B};
(2) \citet{1995A&A...299L...5G}; 
(3) \citet{2000A&A...362L..25G}; 
(4) \citet{1996A&A...309L..35B};
(5) \citet{1999A&A...350L..31G};
(6) \citet{1999A&A...349L..45K};
(7) \citet{2007A&A...462L..49E};
(8) \citet{2008A&A...489..543E};
(9) \citet{2010ApJ...722.1035M}.

\end{deluxetable}

\begin{deluxetable}{lrrrrr}
\scriptsize
\tablecaption{TDE Candidate Radio Results \label{tab:results}}
\tablehead{
\colhead{Source}&  
\colhead{Beam} & 
\colhead{$\sigma$} & 
\colhead{$S_{rad}$} & 
\colhead{$\alpha$} & 
\colhead{$\delta$} \\
 & (arcsec$^2$) &   ($\mu$Jy)  &   ($\mu$Jy)  &   (J2000) & (J2000) \\ 
}
\startdata
IC 3599                 &  $2.0 \times 1.8$ &  14 & $185 \pm 28$ &  12:37:41.19 &  +26:42:27.6 \\ 
RX J1420+5334           &  $2.8 \times 2.0$ &  19 & $114 \pm 24$ & 14:20:25.20 & +53:33:55.0 \\ 
NGC 5905                &  $2.7 \times 2.0$ & 42 & $< 200$ &  \dots & \dots \\ 
RX J1624+7554           &  $3.2 \times 2.0$ & 17 & $< 51$ & \dots & \dots     \\ 
RX J1242-1119           &  $3.9 \times 1.9$ & 18 & $<54$ & \dots & \dots     \\ 
SDSS J132341.97+482701.3 & $3.3 \times 2.1$ & 34 & $<102$ & \dots & \dots       \\ 
SDSS J131122.15-012345.6 & $3.5 \times 1.8$ & 19 & $<57$ & \dots & \dots \\ 
\enddata
\end{deluxetable}

We detected radio sources for three of the seven candidates.  We discuss each of these
sources in detail here.

\subsection{IC 3599}

\begin{figure}
\includegraphics[height=0.8\textheight,angle=-90]{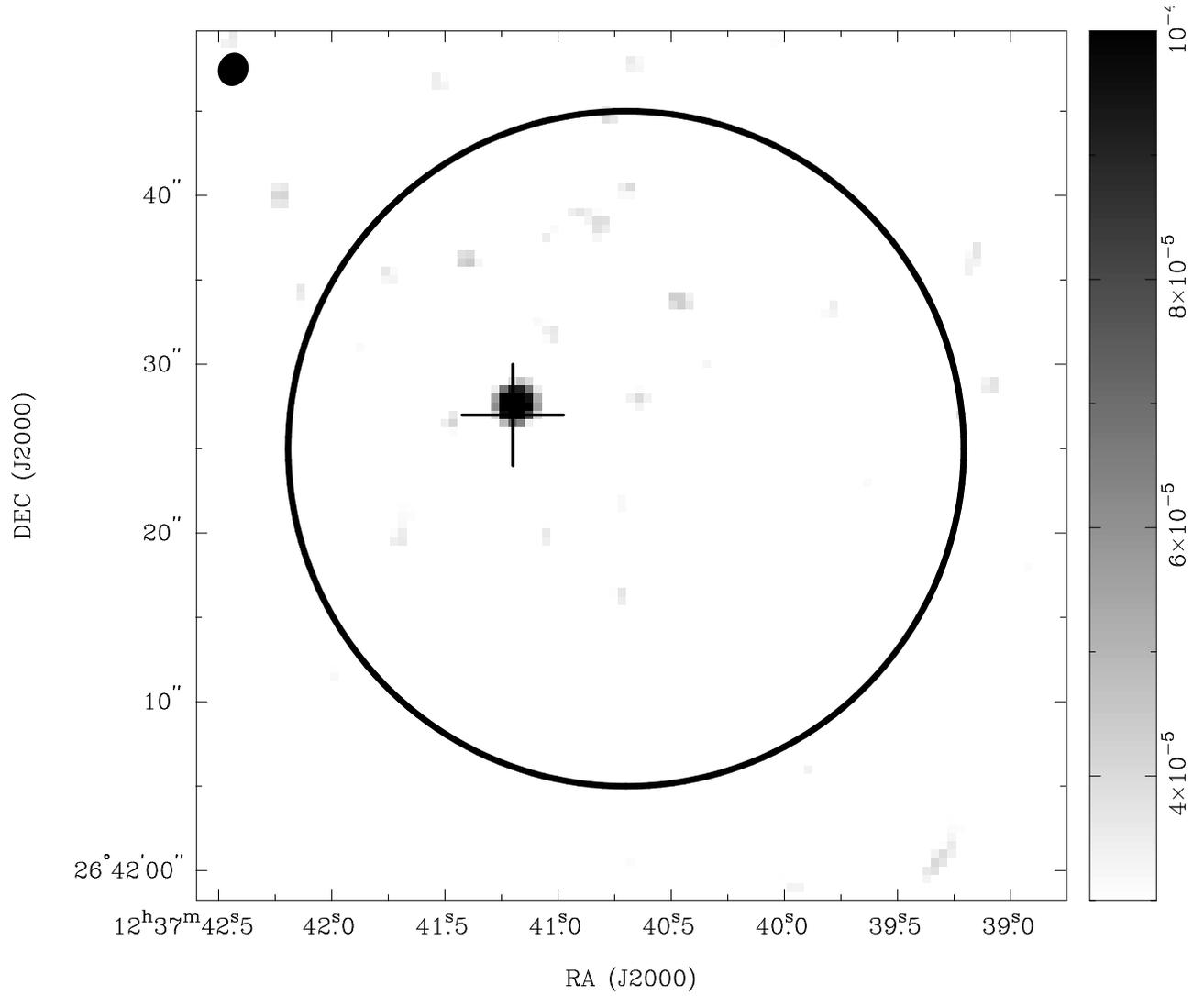}
\caption{VLA image of IC 3599 between 2 and 4 GHz.  
The synthesized beam is show in the upper lefthand corner.
The cross marks the optical centroid for IC 3599.  The large circle
marks the 20\arcsec\ error radius for the ROSAT pointed observation.
The colorbar gives flux density in units of Jy.
\label{fig:ic3599}}
\end{figure}

Figure~\ref{fig:ic3599} shows the image of the radio counterpart for a TDE detected
with ROSAT in 1990.  The image has been made with only baselines longer than $20 k\lambda$ in
order to filter out faint, diffuse emission associated with the galaxy.  The remaining
object is well-fit as a compact point source with flux density of 185$\pm 28\, \mu$Jy. The centroid of
the source is well within the 40\arcsec\ error circle associated with ROSAT and is within
0.2\arcsec\ of the optical nucleus of the galaxy.  We estimate a spectral index from 2 to 4 GHz of
$\beta=-0.3 \pm 0.3$ based on the first order Taylor-series term of the flux density images
of $-22\ \mu$Jy (using $S \propto \nu^\beta$).

\subsection{RX J1420.4+5334}

\begin{figure}
\includegraphics[height=0.8\textheight,angle=-90]{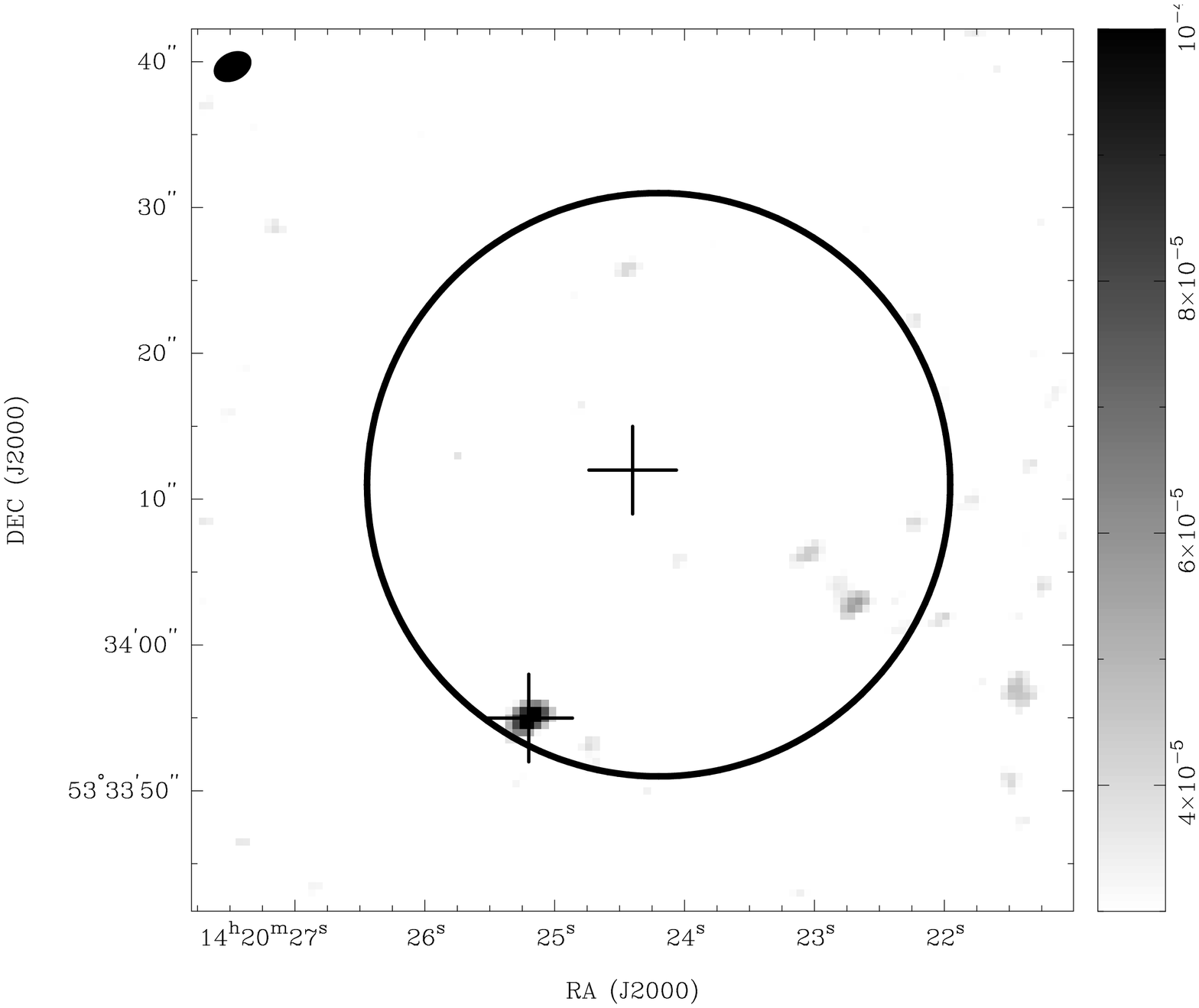}
\caption{VLA image of RX J1420+5334 between 2 and 4 GHz.  
The synthesized beam is show in the upper lefthand corner.  
The central cross marks the position of galaxy A from \citet{2000A&A...362L..25G}.
The other cross marks SDSS J142025.18+533354.9.
The large circle
marks the 20\arcsec\ error radius for the ROSAT pointed observation.
The colorbar gives flux density in units of Jy.
\label{fig:rxj1420}}
\end{figure}

Figure~\ref{fig:rxj1420} shows the image of a potential radio counterpart for a TDE detected
by ROSAT in 1990.  The radio emission of $114 \pm 24\ \mu$Jy
is associated with the galaxy SDSS J142025.18+533354.9
and is offset from the association with galaxy A identified
by \citet{2000A&A...362L..25G} (2MASS J14202436+5334117).  SDSS J142025.18+533354.9 is a galaxy
with a $g$ magnitude of 22.5 and no previously measured redshift. 
We fit a third-order Taylor series expansion of the flux density that leads to $\beta$ changing 
from $-1.2$ to $-0.4$ as the frequency changes from 2 to 4 GHz.  For convenience, we estimate
$\beta \approx -0.8 \pm 0.4$.

We also imaged archival VLA data from experiment AW 651 obtained on 10 and 31 
Mar 2005 at 1.4 GHz in pseudo-continuum mode with 25 MHz of bandwidth
in two IF bands.  These B configuration observations had a resolution of
$5.3 \times 5.3$ arcsec$^2$.  The observations were centered 13.1\arcmin\ 
away from the center of the error circle for RXJ1420+5334, leading to
a primary beam attenuation factor of 0.55.  We find no source within the ROSAT
error circle with a uncorrected RMS of 38 $\mu$Jy.  A primary beam
corrected flux density $3\sigma$ upper limit is then 207 $\mu$Jy.
This flux density upper limit is similar to the value we measure in our 2012 
observations.

We obtained a single 600\,s spectrum of galaxy SDSS\,J142025.18+533354.9
with the Low Resolution Imaging Spectrometer 
\citep[LRIS;][]{occ+95}
mounted on the 10\,m Keck I telescope on 2012 August 
19 UT.  LRIS employs a dichroic beam splitter and was configured with the 
400\,lines\,mm$^{-1}$ / 8500\,\AA\ grating on the red side and the 
600\,lines\,mm$^{-1}$ /4000\,\AA\ grism on the blue side.  Using the
1\arcsec\ slit, this resulted in a resolution of $\sim 6.5$\,\AA\ on the 
red side and $\sim 4.0$\,\AA\ on the blue arm, with a total wavelength
coverage from $\approx 3500$--$10200$\,\AA.

\begin{figure}
\includegraphics[width=\textwidth]{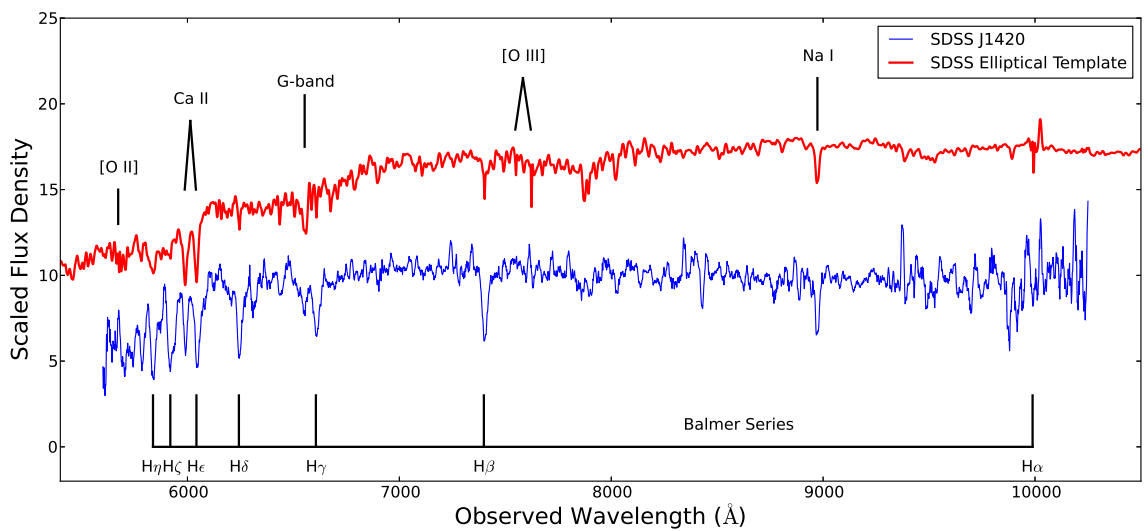}
\caption{Optical spectrum of the galaxy 
SDSS\,J142025.18+533354.9.  
Based on a number
of absorption features, we measure a redshift of $z = 0.522 \pm 0.002$ for
this galaxy.  Shown for comparison is the composite spectrum of early-type
galaxies from the Sloan Digital Sky Survey.  The lack of bright
nebular emission lines (in particular [\ion{O}{2}] and H$\alpha$), together
with the prominent Balmer break, indicate an older stellar population lacking
current star formation.  The lack of broad emission lines (i.e., 
a broad-line region) implies that SDSS\,J142025.18+533354.9
is not a typical active 
galaxy.}
\label{fig:sdssj1420}
\end{figure}

The resulting spectrum of SDSS\,J1420 is plotted in Figure~\ref{fig:sdssj1420}.
Super-posed on a relatively red continuum, we identify a number of 
marginally resolved absorption features corresponding to the Balmer
series (H$\beta$, H$\gamma$, H$\delta$, H$\epsilon$, H$\zeta$, 
and H$\eta$), \ion{Ca}{2} H+K, \ion{Na}{1} D, and the G-band.  All observed
features are consistent with a common redshift of $z = 0.522 \pm 0.002$.  
We find no evidence for any significant emission features over the range
of our spectrum, which covers rest frame lines including [\ion{O}{2}],
[\ion{O}{3}], H$\beta$, and H$\alpha$.
The limit on H$\alpha$ emission is $\sim 5 \times 10^{-16} {\rm\ erg\ cm^2\ s^{-1}}$, 
or an H$\alpha$ luminosity of $< 5 \times 10^{41} {\rm\ erg\ s^{-1}}$ (at $z = 0.522$).  
Using the relation from \citet{1998ARA&A..36..189K}, this corresponds to SFR $< 5\ M_\sun {\rm yr^{-1}}$.

\subsection{NGC 5905}

An extended radio source is found at the position of NGC 5905.   The source has a fitted size of 
$5\arcsec\ \times 4\arcsec$ and a total flux density of 6.7 mJy.  Filtering out baselines shorter
than $80 k\lambda$, we find some residual structure but no convincing evidence for a point source at the location of the optical nucleus.  The residual structure 
has a peak intensity of 200 $\mu$Jy beam$^{-1}$ and is offset from the location of the
optical nucleus. The image rms within the ROSAT error circle is 42 $\mu$Jy.  
We estimate a $3\sigma$ upper limit of 126 $\mu$Jy for a point source
at that location but conservatively assign an upper limit of 200 $\mu$Jy.

\section{Discussion \label{sec:discussion}}

We argue that the radio emission from IC 3599 is associated with the nucleus
responsible for the X-ray TDE candidate.  IC 3599 exhibited variable optical emission lines 
that are consistent with a Seyfert nucleus prior to and after the detection
of the X-ray flare
\citep{1995MNRAS.273L..47B,1995A&A...299L...5G,1999A&A...343..775K}.  
In particular, optical spectroscopy demonstrated an evolution of the
Seyfert-like spectrum on a timescale of $\sim 1$ years from a Sy 1.5 to
Sy 1.9 classificaiton.  The disk of IC 3599 is $\sim 20\arcsec$,
comparable in size to the error circle from the ROSAT observations.  
This spectral evolution and the association with the galactic nucleus
indicate that the X-ray source is likely associated with IC 3599.
Chandra X-ray spectroscopy 12 years after the ROSAT detection found
an X-ray luminosity lower by a factor of $\sim 100$ from the peak and
a spectrum with power-law photon index of $\Gamma=3.6$, significantly
steeper than a typical Seyfert galaxy \citep{2004MNRAS.349L...1V}.
We conclude then that the radio emission is associated with a nucleus
that underwent a significant evolution over the TDE epoch
and that the resulting nucleus does not resemble a typical Seyfert source.

The association of the radio source in the field for RX J1420.4+5334 with the TDE 
event is less certain.  
Galaxy A (2MASS J14202436+5334117) was the brightest optical galaxy in the field 
and, therefore, likely to be the nearest galaxy.  However, optical
spectroscopy after the TDE showed no evidence for AGN like emission lines;
optical photometry found no evidence for nuclear variability.  Thus, the
association of the TDE with galaxy A is not strong.  
Galaxy B identified by \citep{2000A&A...362L..25G} is SDSS J142024.52+533415.7,
is at a distance of 4\arcsec\ from galaxy A, and
has an SDSS spectrum that shows strong emission lines, potentially consistent with
an AGN and is, therefore, also a candidate as host for the X-ray transient.
We note that the galaxy associated with the radio emission, SDSS J142025.18+533354.9,
shows no evidence for broad or narrow emission lines, although emission lines
could be obscured by the accretion region or the host galaxy such as in
an X-ray bright optically normal galaxy \citep{2002ApJ...571..771C}.

What is the probability of an unassociated radio source appearing in the ROSAT error circle
for RX J1420.4+5334?
The differential space density of 1.4 GHz radio sources is $\sim 3 \times 10^{10}\ {\rm Jy^{-1}\ sr^{-1}}$
at 100 $\mu$Jy \citep{1985AJ.....90.1957M}.    For a 20\arcsec\ error radius, we have an expectation of $\sim 0.1$
radio sources above our detection threshold.  Therefore, we cannot conclusively
identify the detected radio source associated with SDSS J142025.18+533354.9 
as the TDE counterpart without further evidence such as variability or a distinguishing radio
spectrum. 

The upper limit on the SFR in RX J1420.4+5334 
translates to a radio synchrotron flux density from star formation of $< 5 \mu$Jy
\citep{2009MNRAS.397.1101G}.  Thus, most of the flux is likely to be associated with 
a compact nucleus.  

\subsection{Relativistic Jet Creation}

\begin{figure}
\includegraphics[width=\textwidth]{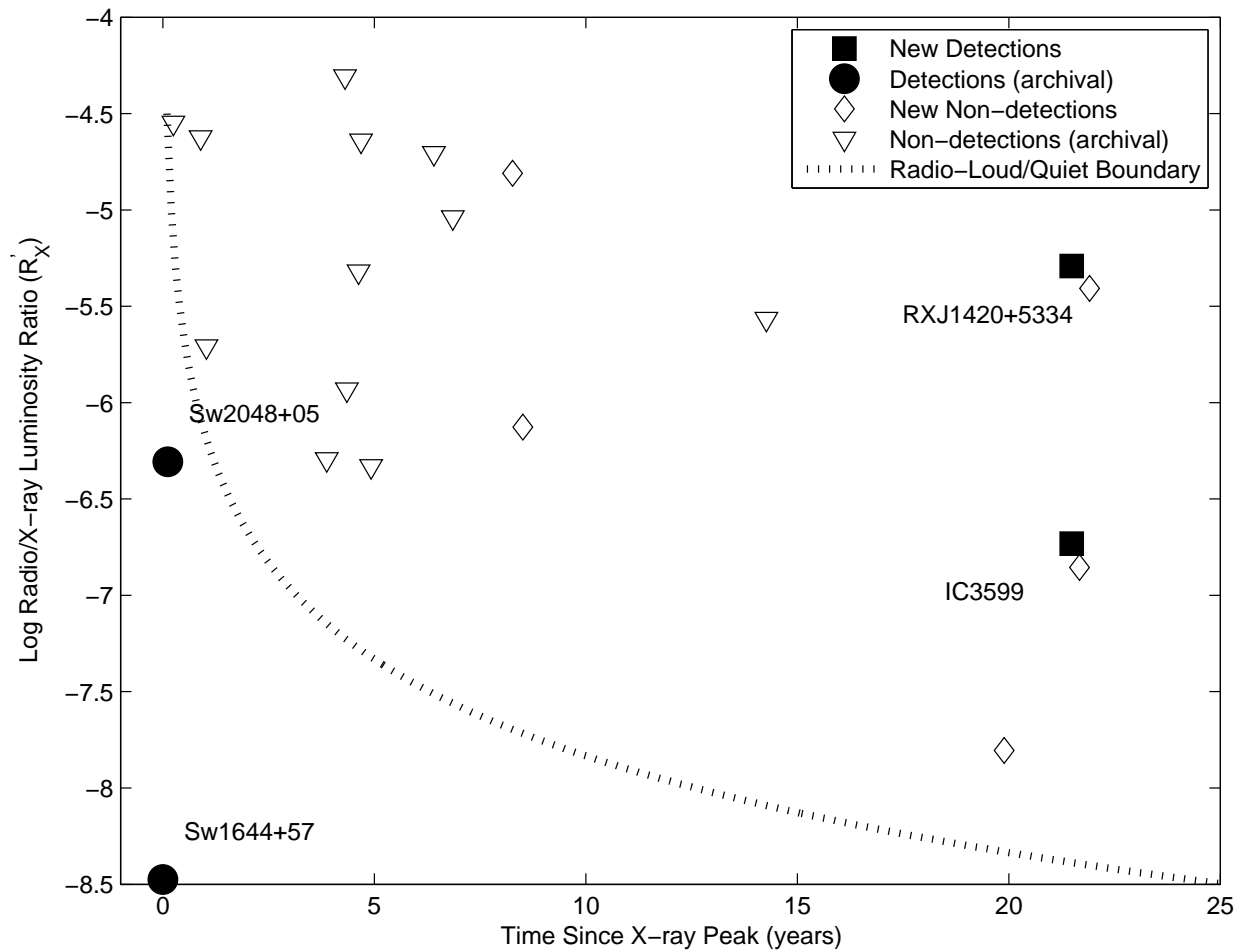}
\caption{Ratio of radio and peak X-ray luminosity as a function of time from X-ray
peak luminosity.  
Solid squares are detections from this paper; solid
circles are detections from the literature; open diamonds are upper limits from this paper;
and open triangles are upper limits from the archival data and the literature.
Detections are labeled.
The dotted curve indicates the expected boundary between radio-loud and radio-quiet 
sources for a solar mass star disrupted by a $10^7 M_\sun$ black hole.
\label{fig:timeradioxray}
}
\end{figure}

We summarize detections and upper limits in Figure~\ref{fig:timeradioxray}.  We show 
the ratio of radio to peak X-ray luminosity.
Data are taken from the new observations obtained in this paper, archival data analyzed in this
paper, upper limits from NVSS and FIRST, an upper limit for the UV-selected source GALEX J141929+525206
\citep{2006ApJ...653L..25G,2011ApJ...732L..12B}, upper limits for the optically-selected source
TDE2 and PTF 10iya \citep{2011ApJ...741...73V,2012MNRAS.420.2684C}, and detections
for the sources Sw 1644+57 and Sw 2048+05 \citep{Bloom08072011,2012ApJ...753...77C}.
For the UV- and optically-selected sources we use estimates of the bolometric luminosity for
the peak X-ray luminosity. 

We define the luminosity ratio
\begin{equation}
R_X^\prime = \log (\nu L_R / L_X^{peak}),.
\end{equation}
where $L_R$ is the spectral luminosity at radio frequency, $\nu$, and $L_X^{peak}$ is the peak X-ray
luminosity.
This ratio is similar to the quantity $R_X$ derived from simultaneous radio and X-ray luminosities
\citep{2003ApJ...583..145T}.  The value $R_X=-4.5$ is a characteristic dividing line between
radio-loud and radio-quiet active galactic nuclei.  Because $L_X$ declines substantially following peak
and will be orders of magnitude lower years or decades later than its peak value, $R_X^\prime$ will 
be smaller than $R_X$ when the radio source is detected.  

As discussed in $\S\ref{sec:intro}$, two models have been proposed for radio emission from TDEs.  In the ``internal'' jet emission model of van Velzen et al.~(2011), the radio emission directly traces the instantaneous accretion rate.  In this case, we can use the theoretical predictions for evolution of the accretion rate as a proxy for $L_X$ as a means of exploring the radio loudness
of these sources.  The accretion rate is expected to scale as $(t/t_{fb})^{-5/3}$ \citep{1988Natur.333..523R,1989IAUS..136..543P}.  
The characteristic fall back time scale ($t_{fb}$)
for a black hole with mass $10^7 M_\sun$ tidally capturing a solar mass star at a periastron
radius equal to the tidal radius is $\sim 0.1$ y \citep[e.g., ][]{2009MNRAS.400.2070S}.  
Detailed modeling for these systems indicates that our assumptions
of black hole and stellar mass are consistent with the X-ray light curves \citep{2011ApJ...736..126M}.
We plot the expected $R_X^\prime$ 
for this case in Figure~\ref{fig:timeradioxray}.  All of our detections and non-detections fall on
the radio-loud side of this boundary with the exception of the very early-time measurements for
Sw 1644+57 and Sw 2048+05.  However, the radio emission in these sources is well fit by the CNM interaction model (see \S\ref{sec:blastwave} below) and is strongly affected by relativistic beaming.  Our two new detections are both 1.5 to 3 orders of magnitude above the radio-loudness boundary.

Thus, we can conclude that our detections indicate that a fraction $f_{late} \gsim 10$\% of TDE 
candidates found at X-ray wavelengths appear to have relativistic jets present at late times.  Further, our non-detections are not sufficiently sensitive to exclude the possibility that $f_{late}$ may be significantly
larger.  At earlier times ($\sim 1$ year), the existing radio limits are also above the radio-loud boundary but less significantly so.  The dearth of detections shortward of $\sim 20$ year suggests, however, that relativistic jets may be launched when the accretion rate drops below a certain threshold.  This is consistent with the burst model proposed by \citet{2011MNRAS.417L..51V}.  For our assumption of a $10^7 M_\sun$ black hole disrupting a solar mass star at the tidal radius, the accretion rate at 20 years 
is $4 \times 10^{-4} M_\sun \, {\rm y^{-1}} \approx 4 \times 10^{-3} \dot{M}_{Edd}$.  
Thus, the accretion rate threshold for jet formation appears to be in the range $10^{-2}$
to $10^{-3} \dot{M}_{Edd}$.

An obvious caveat to the above discussion is that, although we have assumed a fixed fall-back time of $t_{\rm fb} \sim 0.1$ years above, in reality $t_{\rm fb}$ can vary significantly between TDEs.  Another important caveat is that IC 3599 showed signs of Seyfert-like activity in its optical emission lines prior to the TDE \citep{1999A&A...343..775K}.  Thus, IC 3599 may not have been truly quiescent.  Nevertheless, the X-ray variations are consistent with tidal disruption of a star leading to super-Eddington accretion
\citep{2011ApJ...736..126M} and it has been suggested that the TDE rate could be enhanced during AGN phases which follow the merger of a binary SMBH (e.g.,\citealt{2011MNRAS.412...75S,2012arXiv1208.4954J}; \citealt{2012ApJ...749...92K}).  In the microquasar scenario of disk-jet coupling, a pre-existing jet would have been quenched during this phase and, thus, must have been re-launched after the TDE.

\subsection{Blast Wave Model Implications}
\label{sec:blastwave}

In this section we discuss the implications of the radio detections in RX J1420.4+5334 and IC 3599 within the CNM interaction (blast wave) model developed by GM11 and applied to model Swift J1644+57 by MGM12.  An impulsive outflow of energy $E_{\rm j} = E_{52}10^{52}$ ergs becomes non-relativistic at the radius and time at which it sweeps up a CNM mass comparable to its own:
\be
R_{\rm dec} = \left(\frac{3E_{\rm j}}{4\pi n m_{p}c^{2}}\right)^{1/3} \approx 10^{18}{\,\rm cm\,}E_{\rm 52}^{1/3}n_{1}^{-1/3};
\ee
\be 
t_{\rm dec} = R_{\rm dec}/c = 300{\,\rm days\,}E_{\rm 52}^{1/3}n_{1}^{-1/3},
\ee
where $n \equiv n_{1}$ cm$^{-3}$ is the density of the CNM [assumed to be radially-uniform] and we have assumed a spherical outflow since the jet has had sufficient time to spread laterally 
\citep[e.g.,][]{2009ApJ...698.1261Z}.
Note that for realistic values $E_{52} \lesssim 1$ and $n_{1} \gtrsim 1$, transient jets accompanying the X-ray outbursts of either IC 3599 or RX J1420.4+5334 are now well into the non-relativistic Sedov-Taylor phase ($t \approx 22$ years $\gg t_{\rm dec}$).  At times $t \gtrsim t_{\rm dec}$, the [initially relativistic] blastwave velocity and radius scale as $v/c \approx (t/t_{\rm dec})^{-3/5}$ and $r \approx R_{\rm dec}(t/t_{\rm dec})^{2/5}$.  

In the case of RX J1420.4+5334 and IC 3599, the spectral index at the observed frequency $\nu_{\rm obs} \approx 3$ GHz is measured to be $\beta = -0.8\pm 0.4$ and $-0.3\pm 0.3$, respectively.  In the context of a standard synchrotron spectrum generated by power-law electrons (with distribution $dN_{e}/d\gamma_{e} \propto \gamma_{e}^{-p}$ where $p > 2$), this implies that $\nu_{\rm m} < \nu_{\rm obs} < \nu_{\rm cool}$ ($F_{\nu} \propto \nu^{-(p-1)/2}$) in the case of RX J1420.4+5334, while $\nu_{\rm m} < \nu_{\rm obs} < \nu_{\rm cool}$ or $\nu_{\rm sa} < \nu_{\rm obs} < \nu_{\rm m}$ ($F_{\nu} \propto \nu^{1/3}$) in the case of IC 3599, where $\nu_{\rm m}/\nu_{\rm sa}/\nu_{\rm cool}$ are the characteristic, self-absorption, and cooling frequencies, respectively.  Since the latter ordering $\nu_{\rm obs} < \nu_{\rm m}$ is difficult to reconcile with the rapid predicted decline of $\nu_{m} \propto (t/t_{\rm dec})^{-3}$ with time following a single energy injection event, if a flat spectrum $S_\nu \propto \nu^{1/3}$ is indeed found in IC 3599, then this would suggest that core jet emission (van Velzen et al.~2012), or CNM interaction and a very late onset jet ($t_{\rm j} \gg t_{\rm dec}$) may better explain the radio emission.

In order to proceed below we instead assume that $\nu_{\rm m} < \nu_{\rm obs} < \nu_{\rm cool}$ ($F_{\nu} \propto \nu^{-(p-1)/2}$) in both RX J1420.4+5334 and IC3599, in the latter case assuming the lowest physical value $\beta = -0.5$ corresponding to $p = 2$.  This permits us to place constraints on the properties of the jet and the CNM density.  At times $t \gg t_{\rm dec}$ the flux in this frequency range evolves as 
\citep[e.g.,][]{2011Natur.478...82N}:
\begin{eqnarray}
F_{\nu} = 500{\,\rm mJy\,}E_{52}^{0.8-\beta}n_{1}^{0.7+0.5\beta}d_{27}^{-2}\left(\frac{\epsilon_{B}}{0.1}\right)^{0.5-0.5\beta}\left(\frac{\epsilon_{e}}{0.1}\right)^{-2\beta}\left(\frac{\nu}{\rm GHz}\right)^{\beta}\left(\frac{t}{\rm y}\right)^{0.6+3\beta},
\label{eq:Fnu}
\end{eqnarray}
where $d \equiv 10^{27}d_{\rm 27}$ cm is the luminosity distance, $\epsilon_{e}/\epsilon_{B}$ are the fraction of the shock energy placed into the magnetic field and relativistic electrons, respectively.  Now applying equation (\ref{eq:Fnu}) separately to the case of IC3599 ($p = 2$; $\beta \approx -0.5$; $z = 0.02$; $d_{27} = 0.25$)
\begin{eqnarray}
F_{3} = 800{\,\rm mJy\,}E_{52}^{1.3}n_{1}^{0.45}\left(\frac{\epsilon_{B}}{0.1}\right)^{0.75}\left(\frac{\epsilon_{e}}{0.2}\right)\left(\frac{t}{\rm 22\,y}\right)^{-0.9}\,\,{\rm [IC3599]},
\label{eq:FnuIC3599}
\end{eqnarray}
and RX J1420.4+5334 ($p = 3$; $\beta \approx -1$; $z = 0.522$; $d_{27} \approx 9.1$)
\begin{eqnarray}
F_{3} = 0.01{\,\rm mJy\,}E_{52}^{1.8}n_{1}^{0.2}\left(\frac{\epsilon_{B}}{0.1}\right)\left(\frac{\epsilon_{e}}{0.2}\right)^{2}\left(\frac{t}{\rm 22\,y}\right)^{-2.4}\,\,{\rm [RX J1420.4+5334]},
\label{eq:FnuJ1420}
\end{eqnarray}
where in each case we have normalized $E_{52}$, $\epsilon_{e}$, and $n_{1}$ to characteristic values determined from modeling Swift J1644+57 (MGM12; Berger et al.~2012).

Clearly, the fiducial fluxes in equations (\ref{eq:FnuIC3599}) and (\ref{eq:FnuJ1420}) are significantly above or below, respectively, the observed values $F_{3} \approx 0.1-0.2$ mJy.  This discrepency cannot be easily reconciled by invoking a higher or lower CNM density since the dependence of $F_{3}$ on $n$ is relatively weak.  A more plausible reason for this disagreement is the jet energy $E_{\rm j}$ or the strength of the magnetic field behind the shock $\epsilon_{B}$, the latter of which is found to span a wide range $\epsilon_{B} \approx 10^{-4}-0.1$ in normal gamma-ray burst afterglows (the value of $\epsilon_{B}$ was also not well-constrained in the case of Sw J1644+57).  The conclusion of our analysis is that $E_{52}^{1.3}\epsilon_{B} \sim 4\times 10^{-5}$ in the case of IC3599, while $E_{52}\epsilon_{B} \sim 1$ in RX J1420.4+5334.  In IC3599 this implies a weak magnetic field or a much weaker jet energy than that responsible for Swift J1644+57.  The radio emission from RX J1420.4+5334 appears more in line with the expectation of the blast wave model if the magnetic field is in rough equipartition $\epsilon_{B} \sim 1$.  

One prediction of the CNM interaction model is the predicted decay rate of the light curve $F_{3} \propto t^{-\alpha}$ where $\alpha = -0.6$ to $-3$ for $\beta \sim 1-2$, which can be tested by future monitoring of these sources.  Note that in the case of RX J1420.4+5334 the value $\alpha \gtrsim 1.8$ (for $\beta = -0.8$) is only marginally consistent with the NVSS and FIRST upper limits of $F_{3}($t = 5 y$) \approx 2$ mJy.  More accurate measurements of the spectra index, especially in the case of IC3599, will also help distinguish between ongoing core emission versus blast wave models.

\section{Conclusions \label{sec:conclusions}}

We have presented VLA observations of X-ray selected TDEs that indicate one clear
detection of a radio counterpart in IC 3599 and one possible counterpart to
RX J1420.4+5334.  We interpret the radio emission as the result of formation of
a relativistic jet at a time when the accretion rate has dropped substantially
after the TDE.  Our radio non-detections are also consistent with the presence
or absence of a relativistic jet.  Alternatively, we demonstrate that the late-time
emission for RX J1420.4+5334 is consistent with the blast wave model, while the flat spectral index and low radio luminosity in the case of IC 3599 challenge the predictions of the standard blast wave model.  Radio observations that are one to two orders of
magnitude more sensitive than our current results can determine whether a radio-loud
jet is present in the systems with non-detections.

For the two detected sources, further radio observations can demonstrate a closer
relationship to the TDE.  Long-term radio light curves should evolve as 
the mass accretion rate decays.  Radio spectra will be capable of determining the
synchrotron ages of the emission.  And high resolution radio imaging may reveal 
a structure that is consistent with impulsive jet creation in the past two decades.
Such structures could resemble compact symmetric objects as the jet interacts with 
the dense ISM of the galaxy.

\acknowledgements
The National Radio Astronomy Observatory is a facility of the National Science Foundation operated under cooperative agreement by Associated Universities, Inc.  We thank Kelsey Clubb and Alex Filippenko for
assistance with optical spectroscopy.
The authors thank Steve Croft for useful discussions.
Some of the data
utilized herein were obtained at the W. M. Keck Observatory, which is
operated as a scientific partnership among the California Institute of
Technology, the University of California, and the National Aeronautics
and Space Administration (NASA); the observatory was made possible by
the generous financial support of the W. M. Keck Foundation.
S.B.C. acknowledges generous financial assistance from Gary \& Cynthia Bengier, the Richard \& Rhoda Goldman Fund, and US National Science Foundation (NSF) grant AST-0908886.
J. S. B. was partially supported by NASA/NNX10AF93G.


\end{document}